\documentclass[12pt]{article} 
\title{U(N) Based Transformations in N-Squared Dimensions}  
\author{{\it Richard Shurtleff~}\thanks{affiliation and mailing 
address: Department of Sciences, College of Arts and Sciences, 
Wentworth Institute of Technology, 550 Huntington Avenue, 
Boston, MA, USA, ZIP 02115, telephone number: (617) 989-4338, e-mail address: shurtleffr@wit.edu}} 
\begin{document} 
          
\maketitle

\begin{abstract}

Consider the example of the relationship of the group O(3) of rotations in 3-space to the special unitary group SU(2). Given other unitary groups, what transformations can we find? In this paper we describe a method of constructing transformations in an N-squared dimensional manifold based on the properties of the unitary group U(N). With N = 2, the method gives the familiar rotations, boosts, translations and scale transformations of four dimensional spacetime.  In the language of spacetime now applied for any N, we show that rotations preserve distances and times. There is a scale-changing transformation. For N more than 2, boosts do not preserve generalized spacetime intervals, but each N-squared dimensional manifold contains a 3+1 dimensional subspace with invariant spacetime intervals. That each manifold has a four dimensional spacetime subspace correlates with the property of U(N) that U(2) is a subgroup.  
   
\vspace{2cm}



\end{abstract}

\pagebreak

\section{Introduction} \label{intro}

The process builds on the properties of the unitary group U($N$) to produce transformations in an $N^{2}$ dimensional manifold. The
process differs from generalizations of spacetime based on the invariance of various quadratic forms\cite{KK} and from generalizations based on covariance of selected equations, e.g. Maxwell's.\cite{Bateman} We find scale factors, but not the full conformal group. In particular, we do not get the inversions one finds with the covariance of Maxwell's equations.

To illustrate, consider the prototype $N$ = 2. The unitary group U(2) of $2\times 2$ matrices with matrix multiplication can be considered to be a transformation group. Each matrix $U$ can be written as a transformation, $U$ = $\exp{(i \alpha_{\rho} \sigma^{\rho}/2 )},$ where $\rho \in$ $\{1,2,3,4\},$ the $\alpha$s are a $4$-tuple of real numbers, and the generators $\sigma^{\rho}/2$ form a basis of $2\times 2$ hermitian matrices. For example, the $\sigma^{\mu}$s can be the Pauli spin matrices $\sigma^{i},$ $i \in$ $\{1,2,3\},$ with the $2 \times 2$ identity matrix $\sigma^4.$  

With these $\sigma^{\mu}$s, one can build $4 \times 4$ matrices $J^{i}_{(2+2)},$ $K^{j}_{(2+2)}$ and $P^{\mu}_{(2+2)},$ 
\begin{equation} \label{2JKs}
J^{i}_{(2+2)} = \frac{1}{2}\pmatrix{\sigma^{i} && 0 \cr 0 && \sigma^{i}} \quad ; \quad K^{j}_{(2+2)} = \frac{1}{2}\pmatrix{\sigma^{j} && 0 \cr 0 && -\sigma^{j}} \quad ; \end{equation}
\begin{equation} \label{2Ps}
P^{\mu}_{(2+2)} = \frac{1}{2}\pmatrix{0 && \sigma^{\mu} \cr 0 && 0} \quad ,
\end{equation}
which have the commutation relations
$$ [J^{i},J^{j}]  = i \epsilon^{ijk} J^{k}  \quad ; \quad [J^{i},K^{j}]  = i \epsilon^{ijk} K^{k}  \quad ; \quad [K^{i},K^{j}]  =  -i \epsilon^{ijk} J^{k} \quad
$$
\begin{equation} \label{2Pcomm1} [P^{\mu},J^{j}]  =  i \epsilon^{\mu j \sigma} P^{\sigma}  \; ; \; [P^{\mu},K^{j}] = - i (\delta^{\mu j} \delta^{\sigma 4} + \delta^{\sigma j} \delta^{\mu 4}) P^{\sigma} \; ; \;   [P^{\mu},P^{\nu}] = 0 
  \quad .
\end{equation}
We drop the $(2+2)$ subscript because the commutation relations apply also to other reps, $\epsilon$ is the completely antisymmetric symbol with components $\epsilon^{ijk}$ = $(j-i)(k-i)(k-j)/2 $ and $\epsilon^{4jk}$ = 0, and $\delta^{\mu \nu}$ is the identity matrix with ones on the diagonal and zeros everywhere else.

The commutation relations (\ref{2Pcomm1}) signal the Lie algebra of the Poincar\'{e} group of spacetime transformations. The angular momentum matrices $J^{j}$ generate rotations, boosts are generated by the $K^{j}$, and translations are generated by the momentum matrices $P^{\mu}$. From the above commutation relations, a realization of the generators $J^{j}_{(4)}$ and $K^{j}_{(4)}$ for spacetime transformations in $3+1$ dimensions can be read off as the coefficients of the $P^{\mu}$ on the right in (\ref{2Pcomm1}),
\begin{equation} \label{2JK4} 
J^{j}_{(4)\, \mu \nu} =  i \epsilon^{\mu j \nu}  \; ; \; K^{j}_{(4)\, \mu \nu} = - i (\delta^{\mu j} \delta^{\nu 4} + \delta^{\nu j} \delta^{\mu 4})   \quad .
\end{equation}
The $J^{j}_{(4)}$ and $K^{j}_{(4)}$ obey the commutation relations (\ref{2Pcomm1}). The momentum matrices $P^{\mu}_{(4)\, \alpha \beta}$ can be found, but by other means. The transformations $\exp{(i \theta J^{j}_{(4)})}$ are the rotations in 3-space about the $j$ = 1,2,3 axes and $\exp{(i \phi K^{j}_{(4)})}$ are boosts in spacetime along the $j$ = 1,2,3 directions.

The process just described leads from properties of the unitary group U(2) to a representation of rotations and boosts in 4-dimensional spacetime. 

Throughout, with higher dimensional spaces, we use the terminology of the $N$ = 2 case of four-dimensional spacetime, with the understanding that only for $N$ = 2 do the terms retain their original meaning.

Generalizing to any $N \geq$ 2, the process begins by identifying key algebraic conditions of the $J$s, $K$s and $P$s in the $2N$ representation (above, the $2+2$ rep).  Next, the commutation relations of the $J$s, $K$s and $P$s are determined. The coefficients of $P$ for the $[P,J]$s become the generators $J_{(N^2)}$ of `rotations' in $N^{2}$ spacetime. The coefficients of $P$ for the $[P,K]$s become the generators $K_{(N^2)}$ of `boosts'. The momentum matrices $P_{(N^2)}$ can be found by solving the commutation relations with the $J_{(N^2)}$s and $K_{(N^2)}$s which, by this point, are known. Knowing the generators determines the transformations and we are done.

The algebraic conditions are central. There are three, labeled I, II, and III. The first says that, with a rotation, the momentum matrix $P^{\mu}_{rs}$ is required to transform both as a vector with index $\mu$ and as a second rank tensor with indices $r,s.$ This means that, for any angle $\theta_{j},$ i.e. for any $N^2$-tuple of real numbers, we get the same transformed matrix by two ways, (I):
$$
{\mathrm{(tensor)}} \quad [\exp{(-i \theta_{j} J^{j}_{(2N)})}]_{ru}\; P^{\mu}_{uv} \; [\exp{(+i \theta_{j} J^{j}_{(2N)})}]_{vs} = [\exp{(-i \theta_{j} J^{j}_{(N^{2})})}]^{\mu \sigma} \; P^{\sigma}_{rs} \quad {\mathrm{(vector)}} \quad , 
$$
where $\mu \in$ $\{1,...,N^2\},$ $r,s,u,v \in$ $\{1,...,N\}$ and summation over repeated indices is understood. The second says the same thing for boosts, (II):
$$
{\mathrm{(tensor)}} \quad [\exp{(-i \phi_{j} K^{j}_{(2N)})}]_{ru}\; P^{\mu}_{uv} \; [\exp{(+i \phi_{j} K^{j}_{(2N)})}]_{vs} = [\exp{(-i \phi K_{(N^{2})})}]^{\mu \sigma} \; P^{\sigma}_{rs} \quad {\mathrm{(vector)}} \, ,
$$
where the boost parameters $\phi_{j}$  form another $N^2$-tuple of real numbers. By I and II, momentum matrices are both second rank tensors and vectors under rotations and boosts. 

Using the Hausdorff formula and the `CopyCat Theorem' discussed herein, we show that these requirements follow from the commutation rules,
$$ \quad {\mathrm{I.}} \quad [P^{\mu},J^{\nu}] = \rho^{\mu \nu \sigma} P^{\sigma} \quad {\mathrm{and}} \quad  {\mathrm{II.}} \quad [P^{\mu},K^{\nu}] = \kappa^{\mu \nu \sigma} P^{\sigma} \quad ,
$$ 
so the commutators with momentum matrices must be sums of momentum matrices. Because the indices $\mu \nu \sigma$ on the coefficients $\rho$ and $\kappa$ run from 1 to $N^2,$ it is convenient to extend the formalism to include $N^2$ $J$s and $K$s, so the index $\mu$ for the matrix generators $J^{\mu}$s and $K^{\mu}$s runs from 1 to $N^2.$

I and II so far treat $J$ and $K$ algebraically equally.  Boosts and rotations need to be distinguished one from the other. To distinguish angular momentum $J$ from boost generators $K,$ one requires that the coefficients $\rho^{\mu \nu \sigma}$ must be antisymmetric in $\mu \nu$ while the $\kappa^{\mu \nu \sigma}$s must be symmetric in $\mu \nu.$ Amend I and II to include these symmetries. 

Next, include properties of U($N$). Consider the group U($N$) of unitary $N \times N$ matrices combined with matrix multiplication . Since each matrix $U$ is unitary, the generators $h^{\mu}_{(N)}$ of U($N$) are hermitian, i.e. $U$ = $\exp{(i \alpha h_{(N)})}$ for some choice of real $\alpha$s. Let $h^{\mu}_{(N)}$ be a basis of $N \times N$ hermitian matrices. 

The third requirement is: III. There exists a representation in which the matrices $J^{\mu}$, $K^{\mu}$ and $P^{\mu}$ are built solely from the basis matrix $h^{\mu}_{(N)},$ {\it{for each $\mu.$}}  Requirement III ensures that rotations, boosts, and translations have properties that can be traced back to the properties of $N \times N$ hermitian matrices. 

`Built from' means that each $J^{\mu}$, $K^{\mu}$ and $P^{\mu}$ can be written in block matrix form with a numerical multiple, possibly zero, of $h^{\mu}_{(N)}$ in each block. We see below that the $J^{\mu}$, $K^{\mu}$ and $P^{\mu}$ are built with $h$s in the form of a $2N$ dimensional rep of `block-matrices', $2N \times 2N$ matrices with components arranged as multiples of the $N \times N$ matrices $h^{\mu}.$

In Sec. 2, some properties of U($N$) and a hermitian basis of generators $h^{\mu}$ are discussed. Sec. 3 applies the algebraic conditions I, II, III to find commutation relations for angular momentum $J,$ boost $K,$ and momentum $P,$ the generators of rotations, boosts and translations, respectively. The commutation relations found in Sec. 3 for $N >$ 2 differ from the Poincar\'{e} commutation relations for $N$ = 2 because the coefficients differ and the number of dimensions differ. Sec. 4 finds a representation of the angular momentum and boost generators for $N^2$-dimensional spacetime. Sec. 5 shows that momentum matrices $P$ are arrays of Clebsch-Gordan coefficients in reducible reps. Sec. 6 defines specific matrices, `the utility rep', and shows that distances in $N^2-1$ dimensional space as well as times are invariants of rotations. Boosts have no general coordinate quadratic invariants like rotations have, but each $N^2$-dimensional spacetime has a four-dimensional subspace with the usual quadratic invariants of $3+1$ spacetime transformations. Sec. 7 considers four-dimensional spacetime as the special case $N$ = 2. An Appendix proves a theorem needed in Sec. 6 to find momentum matrices.  

We find that rotations in $N^2$ dimensions are related to the commutators of the $h^{\mu}_{(N)}$s, the boosts are related to the {\it{anti}}commutators of the $h^{\mu}_{(N)}$s and the translations are arrays of Clebsch-Gordan coefficients that connect various representations of the algebra. Scale transformations occur as time-directed boosts. Rotations, boosts and translations make up the Poincar\'{e} group in spacetime ($N$ = 2). Weyl studied scale transformations \cite{Weyl} so we call the resulting group the `$N$-Poincar\'{e}-Weyl group in $N^2$-dimensional spacetime'.

\section{U(N) and its Generators} \label{FunRep}

Consider the set $W$ of $N\times N$ matrices with complex components. Any matrix $A \in$ $W$ is the sum of a hermitian matrix $(A + A^{\dagger})/2$ and an antihermitian matrix, $(A - A^{\dagger})/2,$
\begin{equation} \label{A}
A = \frac{A + A^{\dagger}}{2} + \frac{A - A^{\dagger}}{2} \quad ,
\end{equation}
where the dagger $\dagger$ denotes the hermitian conjugate, the operation producing the transpose of the complex conjugate, $A^{\dagger}_{rs}$ = $A_{sr}^{\ast},$ $r,s \in$ $\{1,...,N\}.$

Let the set of hermitian $N\times N$ matrices $h^{\mu}_{(N)} \in$ $W,$  $\mu \in$ $\{1,...,N^{2}\}$ form a basis of such matrices.  Then any hermitian matrix $B,$ $B^{\dagger}$ = $B,$  and any anti-hermitian matrix $C,$ $C^{\dagger}$ = $-C,$  can be written as 
\begin{equation} \label{basis}
B = {\beta_{\sigma} h^{\sigma}_{(N)}} \quad ; \quad  C = {i \gamma_{\rho}  h^{\rho}_{(N)}} \quad ,
\end{equation}
for $N^{2}$ real-valued coefficients $\beta_{\mu}$ and $N^{2}$ real-valued coefficients $\gamma_{\mu}.$ The rule of summing over repeated indices is enforced. The first in (\ref{basis}) reflects the intended meaning of `basis', while the second follows from the first because $iC$ is hermitian. 

A unitary matrix $U \in$ $W,$ $U^{\dagger}U$ = $UU^{\dagger}$ = $\mathbf{1},$ can be written as the matrix exponential of an anti-hermitian matrix ${C},$ i.e. $U$ = $\exp{{C}} $ and ${C}^{\dagger}_{rs}$ = $-{C}_{rs}.$ One can confirm this quickly. Consider the formula $\exp{{C}} $ = $ \lim(\mathbf{1}+ C/n)^n$ for $n \rightarrow \infty.$ For infinitesimal $\delta c$ = $C/n,$ 
$$\mathbf{1} = UU^{\dagger}  = (\mathbf{1}+ \delta c)^{n}(\mathbf{1}+ \delta c^{\dagger})^{n} = [(\mathbf{1}+ \delta c)(\mathbf{1}+ \delta c^{\dagger})]^{n} = (\mathbf{1}+ \delta c+ \delta c^{\dagger})^{n} \quad, 
$$ 
so $\delta c+ \delta c^{\dagger}$ = 0, $\delta c$ is antihermitian and $C$ is antihermitian. 

Since $C$ is antihermitian and $U$= $\exp{{C}} ,$ it follows from (\ref{basis}) that, for some real-valued parameters $\gamma_{\mu},$  we have
\begin{equation} \label{U}
U(\gamma)  =  \exp{(i\gamma_{\sigma}h^{\sigma}_{(N)})} \quad .
\end{equation}
The unitary matrices $U \in$ $W$ combined with matrix multiplication as the group multiplication operation form a group, U($N$), and the basis matrices $h^{\mu}_{(N)}$ form a set of generators of that group. 

The commutators and anticommutators of  hermitian matrices are antihermitian and hermitian, respectively. Thus, by (\ref{basis}), we have two sets of real coefficients, $f^{\mu \nu \sigma}$ and $d^{\mu \nu \sigma},$ 
\begin{equation} \label{Comm1} [h^{\mu}_{(N)},h^{\nu}_{(N)}]  = i f^{\mu \nu \sigma} h^{\sigma}_{(N)}  \quad {\mathrm{and}} \quad  \{h^{\mu}_{(N)},h^{\nu}_{(N)}\}  =  d^{\mu \nu \sigma} h^{\sigma}_{(N)}\quad,
 \end{equation}
 where the commutator is defined by $[h^{\mu},h^{\nu}] \equiv$ $h^{\mu}h^{\nu}-h^{\nu}h^{\mu},$ the anticommutator is defined by $\{h^{\mu},h^{\nu}\} \equiv$ $h^{\mu}h^{\nu}+h^{\nu}h^{\mu},$ and matrix multiplication is understood. 

Given the basis $h^{\mu}_{(N)} \in$ $W,$ there is a second basis  $\bar{h}^{\mu}_{(\bar{N})} \in$ $W$ of hermitian matrices called the `anti-rep',
\begin{equation} \label{antih}
 \bar{h}^{\mu}_{(\bar{N}) rs} \equiv -h^{\mu}_{(N) sr} = -h^{\mu \, \ast}_{(N) rs}\quad {\mathrm{(anti\,rep)}} \quad ,
\end{equation}
where the components are labeled with $r,s \in$ $\{1,...,N\}.$ The matrix $\bar{h}^{\mu}_{(\bar{N})}$ is the negative transpose of the $h^{\mu}_{(N)}$ matrix which is also the negative complex conjugate of the $h^{\mu}_{(N)}$ matrix. 

It follows that the commutators and anticommutators of the anti-rep $\bar{h}^{\mu}_{(\bar{N})}$s are 
\begin{equation} \label{Comm1anti} [\bar{h}^{\mu}_{(\bar{N})},\bar{h}^{\nu}_{(\bar{N})}]  = i f^{\mu \nu \sigma} \bar{h}^{\sigma}_{(\bar{N})}  \quad {\mathrm{and}} \quad \{\bar{h}^{\mu}_{(\bar{N})},\bar{h}^{\nu}_{(\bar{N})}\}  = - d^{\mu \nu \sigma} \bar{h}^{\sigma}_{(\bar{N})}\quad,
 \end{equation}
which differs from (\ref{Comm1}) by the sign of the coefficients $d^{\mu \nu \sigma}.$

\section{$N$-Poincar\'{e}-Weyl Commutation Relations} \label{PCR}

The commutation relations involve the `angular momentum matrices' $J^{\mu},$ the  `boost matrices' $K$ and the `momentum matrices' $P$.  Angular momentum matrices generate `rotations', boost matrices generate `boosts', and momentum matrices generate `translations'. These names are used for convenience and are not necessarily indicative of properties of either the generators or the transformations.

We determine the commutation relations by demanding certain characteristics: I. The commutation relations $[P^{\mu},J^{\nu}]$ must be antisymmetric in $\mu \nu$ and be sums of momentum matrices. II. The commutation relation $[P^{\mu},K^{\nu}]$ must be symmetric in $\mu \nu$  and be sums of momentum matrices. Then we demand that
\begin{equation} \label{DemandsIandII} \quad {\mathrm{I.}} \quad [P^{\mu},J^{\nu}] = \rho^{\mu \nu \sigma} P^{\sigma} \quad {\mathrm{and}} \quad  {\mathrm{II.}} \quad [P^{\mu},K^{\nu}] = \kappa^{\mu \nu \sigma} P^{\sigma} \quad ,
 \end{equation}
where we require antisymmetry in $\mu \nu$ for the complex numbers $\rho$ and symmetry for the $\kappa$s,
$$ \rho^{\mu \nu \lambda} = -\rho^{\nu \mu \lambda} \quad {\mathrm{and}} \quad \kappa^{\mu \nu \lambda} = +\kappa^{\nu \mu \lambda} \quad .$$
III. There must be at least one representation that has the matrices $J^{\mu},$ $K^{\mu}$ and $P^{\mu}$ built from the basis matrix $h^{\mu}_{(N)}.$ Each matrix $J^{\mu},$ $K^{\mu}$ and $P^{\mu}$ must be in block-matrix form with each block either a block of zeros or a multiple of $h^{\mu}_{(N)},$ for each $\mu.$

In order to build $J$s, $K$s and $P$s using the $h^{\mu}_{(N)}$s, we note that the symmetry of $[P^{\mu},K^{\nu}]$ in $\mu \nu$ is somewhat tricky because $[h^{\mu},h^{\nu}]$ is antisymmetric in $\mu \nu.$ One way to accomplish this trick is to double the number of dimensions and introduce a strategically placed minus sign with the $K$s. The following definitions of $J^{\mu}_{(2N)},$ $K^{\mu}_{(2N)}$ and $P^{\mu}_{(2N)}$ produce the desired characteristics I, II, and III,
 \begin{equation} \label{anti1} J^{\mu}_{(2N)} \equiv \pmatrix{h^{\mu}_{(N)} && 0 \cr 0 && h^{\mu}_{(N)}}  \quad ; \quad K^{\mu}_{(2N)} \equiv \pmatrix{+ih^{\mu}_{(N)} && 0 \cr 0 && -ih^{\mu}_{(N)}} \quad,
 \end{equation}
 \begin{equation} \label{anti2} P^{\mu}_{(2N)} \equiv  \pmatrix{0 && c_{+}h^{\mu}_{(N)} \cr c_{-}h^{\mu}_{(N)} && 0} \quad ,
 \end{equation}
where $c_{+}$ and $c_{-}$ are complex constants. 

It follows that
 \begin{equation} \label{Comm4} [P^{\mu}_{(2N)},K^{\nu}_{(2N)}] =  \pmatrix{0&&-ic_{+}\{h^{\mu}_{(N)},h^{\nu}_{(N)}\} \cr +ic_{-}\{h^{\mu}_{(N)},h^{\nu}_{(N)}\}  &&0} = \quad
 \end{equation}
\vspace{0.3cm}
$$ = i d^{\mu \nu \sigma}\pmatrix{0&&-c_{+}h^{\sigma}_{(N)} \cr +c_{-}h^{\sigma}_{(N)}  &&0} \quad . $$  
One sees that $[P^{\mu},K^{\nu}]$ is symmetric in $\mu \nu$ by the definition of the coefficients $d^{\mu \nu \sigma}$ in (\ref{Comm1}).

By II, we require the commutator $[P^{\mu},K^{\nu}]$ to be a sum of $P^{\nu}$s. Comparing (\ref{Comm4}) with the definition of the $P^{\mu}_{(2N)}$s in (\ref{anti2}), one sees a sign difference in the off-diagonal blocks which requires that 
$$c_{+} = 0 \quad {\mathrm{or}} \quad c_{-} = 0 \quad , 
$$
so one of the off-diagonal blocks of the $P^{\mu}_{(2N)}$ vanishes.

Since $c_{+}$ = 0 or $c_{-}$ = 0, we define two sets of momentum matrices $P^{(\epsilon_{P})\,\mu}_{(2N)}$ distinguished by the label $\epsilon_{P}$ = $\pm 1,$
\begin{equation} \label{P+P-}  P^{(+)\,\mu}_{(2N)} \equiv \pmatrix{0 && c_{+}h^{\mu}_{(N)} \cr 0 && 0}  \quad ; \quad P^{(-)\,\mu}_{(2N)} \equiv \pmatrix{0 && 0 \cr c_{-}h^{\mu}_{(N)}  && 0} \quad .
 \end{equation}
 Clearly the matrices in each set commute,
\begin{equation} \label{PP}[P^{(+)\,\mu}_{(2N)},P^{(+)\,\nu}_{(2N)}] = 0 \quad {\mathrm{and}} \quad [P^{(-)\,\mu}_{(2N)},P^{(-)\,\nu}_{(2N)}] = 0 \quad.
 \end{equation}
 Translations, the group generated by the momentum matrices, form an abelian subgroup.

{\it{Remark 3.1.}} Due to the simplicity of the $d^{\mu \nu \sigma}$ in U(2), one can arrange for both off-diagonal blocks of $P^{\mu}$ to be nonzero. Since such matrices may not commute, they are called vector matrices. For example, in an often displayed rep, the Dirac gamma matrices are vector matrices with both off-diagonal blocks nonzero.\cite{DIRAC}

One can check that the matrices $J^{\mu}_{(2N)},$ $K^{\mu}_{(2N)}$ and $P^{\mu}_{(2N)}$ satisfy the demands I, II and III. Collecting the commutation relations of the matrices $J_{(2N)}$, $K_{(2N)}$, and $P^{(\epsilon_{P})}_{(2N)}$ we have 
$$ [J^{\mu},J^{\nu}]  = i f^{\mu \nu \sigma} J^{\sigma}  \quad ; \quad [J^{\mu},K^{\nu}]  = i f^{\mu \nu \sigma} K^{\sigma}  \quad ; \quad [K^{\mu},K^{\nu}]  =  -i f^{\mu \nu \sigma} J^{\sigma} \quad
$$
\begin{equation} \label{Pcomm1} [P^{(\epsilon_{P})\,\mu},J^{\nu}]  =  i f^{\mu \nu \sigma} P^{(\epsilon_{P})\,\sigma}  \; ; \; [P^{(\epsilon_{P})\,\mu},K^{\nu}] = -\epsilon_{P} i d^{\mu \nu \sigma} P^{(\epsilon_{P})\,\sigma}  \; ; \;   [P^{(\epsilon_{P})\,\mu},P^{(\epsilon_{P})\,\nu}] = 0 
  \quad ,
\end{equation}
for $\mu, \nu, \sigma \in$ $\{1,...,N^2\}$ and $\epsilon_{P}$ = $\pm 1.$ Some of these are (\ref{Comm4}) and (\ref{PP}), while the rest follow from (\ref{Comm1}).

The commutation relations (\ref{Pcomm1}) make up the $N$-Poincar\'{e}-Weyl commutation relations. Any matrices satisfying the commutation relations (\ref{Pcomm1}) form a representation of the $N$-Poincar\'{e}-Weyl algebra. For $N >$ 2 the commutation relations (\ref{Pcomm1}) differ from the Poincar\'{e} algebra because the coefficients $f$ and $d$ derive from U(3), U(4),..., not U(2), and also the algebra here differs because the range of the indices, 9, 16,..., differs from the four dimensions associated with the Poincar\'{e} algebra.

\section{Fundamental Representation} \label{FunRep1}

In this section, by proving a `CopyCat Theorem', we show that the coefficients $if^{\mu \lambda \nu}$ and $id^{\mu \lambda \nu}$ are $N^2$-dimensional matrices $J^{\lambda}_{(N^2)\, {\mu \nu}}$ and $K^{\lambda}_{(N^2)\, {\mu \nu}}$ satisfying the `Lorentz-Weyl' subalgebra of the $N$-Poincar\'{e}-Weyl algebra. The $J^{\lambda}_{(N^2)}$ and $K^{\lambda}_{(N^2)}$ matrices generate rotations and boosts in $N^2-$dimensional spacetime, which is a definition of `rotation' and `boosts', not an indication of their properties.

Consider the $N$-Poincar\'{e}-Weyl commutation relations (\ref{Pcomm1}). Two are in the form $[P,A]$ = $\alpha P,$ the ones for $A$ = $J$ and $A$ = $K.$ One can prove a `CopyCat Theorem'. Given 
\begin{equation} \label{VABC} [P^{\mu},A^{\lambda}] = a^{\mu \lambda \nu} P^{\nu} \quad {\mathrm{and}} \quad  [P^{\mu},B^{\lambda}] = b^{\mu \lambda \nu} P^{\nu}  \quad,
\end{equation}
one can show that
$$[P^{\rho},[A^{\mu},B^{\nu}]] =     \left( a^{\rho \mu \sigma} b^{\sigma \nu \tau} -  b^{\rho \nu \sigma} a^{\sigma \mu \tau} \right) P^{\tau}  \quad.
$$
If we are also given that $[A^{\mu},B^{\nu}]$ = $s^{\mu \nu \sigma} C^{\sigma}$ and  $[P^{\mu}_{(2N)},C^{\lambda}]$ = $c^{\mu \lambda \nu} P^{\nu}_{(2N)}$  we get
$$   [a^{\mu},b^{\nu}] = s^{\mu \nu \sigma} c^{\sigma}  \quad ,
$$
where $a^{\lambda},$ $b^{\lambda},$ and $c^{\lambda}$ are the matrices 
\begin{equation} \label{ab} a^{\lambda}_{\mu \nu} = a^{\mu \lambda \nu}  \quad {\mathrm{;}} \quad  b^{\lambda}_{\mu \nu} = b^{\mu \lambda \nu}  \quad {\mathrm{;}} \quad  c^{\lambda}_{\mu \nu} = c^{\mu \lambda \nu}  \quad.
\end{equation}
The calculation depends on the linear independence of the $h^{\mu}_{(N)}$s which appear in the off-diagonal block form of $P^{\mu}_{(2N)}$ (\ref{anti2}), so it is important to calculate in the $(2N)$-rep. Thus we have shown the CopyCat Theorem: {\it{the coefficients $a,$ $b$ and $c$ form matrices that obey the same commutation relations as $A,$ $B$ and $C.$}} 

According to the CopyCat Theorem,  the coefficients of the $P^{(\epsilon_{P})\,\sigma}$s on the right sides of the $[P^{(\epsilon_{P})\,\mu},J^{\lambda}]$ and $[P^{(\epsilon_{P})\,\mu},K^{\lambda}]$ commutation relations (\ref{Pcomm1}) form  $N^2$-dimensional matrices $J^{\lambda}_{(N^2)}$ and $K^{\lambda}_{(N^2)},$  
\begin{equation} \label{JKN2} 
{J^{\lambda}_{(N^2)\, {\mu \nu}}}  \equiv  i f^{\mu \lambda \nu} \quad ; \quad {K^{(\epsilon_{P}) \, \lambda}_{(N^2)\, \mu\nu}}  \equiv  -\epsilon_{P} i d^{\mu \lambda \nu} \quad.
\end{equation}
With $A,B,C$ = $J,J,J$ then $A,B,C$ = $J,K,K$ and finally $A,B,C$ = $K,K,J,$ it follows from the $N$-Poincar\'{e}-Weyl algebra (\ref{Pcomm1}) that the matrices  $J^{\mu}_{(N^2)}$ and  $K^{(\epsilon_{P}) \, \nu}_{(N^2)}$ satisfy the following commutation relations,
\begin{equation} \label{Lorentz}
 [J^{\mu},J^{\nu}]  = i f^{\mu \nu \sigma} J^{\sigma}  \quad ; \quad [J^{\mu},K^{(\epsilon_{P}) \, \nu}]  = i f^{\mu \nu \sigma} K^{(\epsilon_{P}) \, \sigma}  \quad ; \quad [K^{(\epsilon_{P}) \, \mu},K^{(\epsilon_{P}) \, \nu}]  =  -i f^{\mu \nu \sigma} J^{\sigma} \quad ,
\end{equation}
which are the $N$-{\it{Lorentz}}-Weyl commutation relations, a subset of the $N$-Poincar\'{e}-Weyl algebra (\ref{Pcomm1}). 

The requirements I and II in Sec. \ref{PCR} can be understood as constraining the transformations of momentum matrices. Consider requirement I: $[P^{\mu},J^{\nu}] = \rho^{\mu \nu \sigma} P^{\sigma}$ which turned into $[P^{(\epsilon_{P})\,\mu},J^{\nu}]  =  i f^{\mu \nu \sigma} P^{(\epsilon_{P})\,\sigma},$ by the time we got to (\ref{Pcomm1}). By the Hausdorff formula
$$ e^{-Y} X e^{+Y} = X + [X,Y] + \frac{1}{2} [[X,Y],Y] + ... + \frac{1}{n!} [...[[X,Y],Y],...,Y] + ... \quad , $$
with matrix $Y$ = $i\theta_{\sigma}J^{\sigma}$ and $X$ = $P^{(\epsilon_{P})\,\mu} $ for requirement I, we get
$$ e^{-i\theta_{\sigma}J^{\sigma}} P^{(\epsilon_{P})\,\mu} e^{+i\theta_{\sigma}J^{\sigma}} = P^{(\epsilon_{P})\,\mu} + i\theta_{\sigma}(i f^{\mu \sigma \rho} P^{(\epsilon_{P})\,\rho}) + \frac{1}{2} i^2\theta_{\bar{\sigma}}\theta_{\sigma}(i f^{\mu \bar{\sigma} \bar{\rho}}i f^{\bar{\rho} {\sigma} {\rho}} P^{(\epsilon_{P})\,{\rho}}) + ... \quad , $$
$$ [e^{-i\theta_{\sigma}J^{\sigma}}]_{r\bar{s}} P^{(\epsilon_{P})\,\mu}_{\bar{s}\bar{r}} [e^{+i\theta_{\sigma}J^{\sigma}}]_{\bar{r}s} = \left(\delta^{\mu}_{\rho}  + i\theta_{\sigma}J^{\sigma}_{(N^2)\,\mu \rho}  + \frac{1}{2} i\theta_{\bar{\sigma}}J^{\bar{\sigma}}_{(N^2)\,\mu \bar{\rho}}i\theta_{\sigma}J^{\sigma}_{(N^2)\,\bar{\rho} \rho} +... \right) P^{(\epsilon_{P})\,{\rho}}_{rs} \quad , $$
and that implies
\begin{equation} \label{tensorVECTOR} {\mathrm{(tensor)}} \quad [e^{-i\theta_{\sigma}J^{\sigma}}]_{r\bar{s}} P^{(\epsilon_{P})\,\mu}_{\bar{s}\bar{r}} [e^{+i\theta_{\sigma}J^{\sigma}}]_{\bar{r}s} = [e^{+i\theta_{\sigma}J^{\sigma}_{(N^2)}}]_{\mu \rho} P^{(\epsilon_{P})\,{\rho}}_{rs} \quad {\mathrm{(vector)}} \quad , 
\end{equation}
where we have used (\ref{JKN2}). 

One sees that $P^{(\epsilon_{P})\,{\mu}}_{rs}$ transforms as a second rank tensor under the rotation matrix generated with the $J^{\sigma}$s. And $P^{(\epsilon_{P})\,{\mu}}_{rs}$ transforms as a vector under the rotation matrix generated with the $J^{\lambda}_{(N^2)}$s. Thus requirements I and II along with the CopyCat Theorem imply that momentum matrices transform both as second rank tensors and as vectors under rotations and boosts.

\section{Momentum Matrices $P$} \label{MomMat}

In this section, we show that momentum matrices are arrays of Clebsch-Gordon coefficients relating certain reps of the $N$-Lorentz-Weyl algebra (\ref{Lorentz}). 

To begin, we consider a way to get representations of the $N$-Lorentz-Weyl algebra (\ref{Lorentz}). Let $J^{\mu}_{(A)\, a a_1}$ and $J^{\nu}_{(B)\, b b_1}$ generate two irreducible reps of U($N$), which means the matrices obey the same  commutation relations (\ref{Comm1}) as the matrices $h^{\mu}_{(N)}$ of the basis rep. In general, the {\it{anti}}commutation relations  (\ref{Comm1}) are not obeyed by the $J_{(A)}$s and $J_{(B)}$s. Now combine the two reps as follows,
\begin{equation} \label{JAB}
	J^{\mu}_{(A,B)rs} = J^{\mu}_{(A,B)a_{1}b_{1},a_{2}b_{2}} \equiv J^{\mu}_{(A)a_{1}a_{2}}\delta^{b_{1}}_{b_{2}} + \delta^{a_{1}}_{a_{2}}J^{\mu}_{(B)b_{1}b_{2}} \quad
\end{equation}
and also
\begin{equation} \label{KAB}
	K^{\mu}_{(A,B)rs} = K^{\mu}_{(A,B)a_{1}b_{1},a_{2}b_{2}} \equiv -i \left(J^{\mu}_{(A)a_{1}a_{2}}\delta^{b_{1}}_{b_{2}} - \delta^{a_{1}}_{a_{2}}J^{\mu}_{(B)b_{1}b_{2}}\right) \quad,
\end{equation}
where $\delta^{a_{1}}_{a_{2}}$ is unity for $a_1$ = $a_2$ and zero otherwise. 

In these expressions the single indices $r$ and $s$ on the left are identified with a prescribed ordering of the double indices $(a_{1},b_{1})$ and $(a_2,b_2),$ respectively, on the right, 
$$r \rightarrow (a_{1},b_{1}) \quad {\mathrm{and}} \quad s \rightarrow (a_{2},b_{2}) \quad .
$$
Thus the matrices ${J^{\mu}_{(A,B)}}$ and ${K^{\mu}_{(A,B)}}$ are square matrices with a dimension equal to the product of the dimensions of reps $A$ and $B.$ 

Based on the fact that $J^{\mu}_{(A)}$ and $J^{\mu}_{(B)}$ each satisfy the fundamental $[h^{\mu}_{(N)},h^{\nu}_{(N)}]$ commutation relations (\ref{Comm1}), e.g. for the $A$-rep: $[J^{\mu}_{(A)},J^{\nu}_{(A)}]$ = $i f^{\mu \nu \sigma} J^{\sigma}_{(A)}  $ and similarly for the $B$-rep, one can show that $J^{\mu}_{(A,B)} $ and $ K^{\mu}_{(A,B)}$ obey the $N$-Lorentz-Weyl commutation relations (\ref{Lorentz}). 

One case is especially important, the $(N^2)$-rep (\ref{JKN2}): For $\epsilon_{P}$ = $+1,$ the $(N^2)$-rep is equivalent to the $(A,B)$ = $(N,\bar{N})$ rep and, for $\epsilon_{P}$ = $-1,$ the $(N^2)$-rep is equivalent to the $(A,B)$ = $(\bar{N},N)$ rep. This means that there exists a similarity transformation $S$ such that
\begin{equation} \label{JN2}
	S^{\mu}_{\sigma}J^{\lambda}_{(N^2)\sigma \nu} = \left(h^{\lambda}_{(N)\mu_{1} \rho_{1}}\delta^{\mu_{0}}_{\rho_{0}} + \delta^{\mu_{1}}_{\rho_{1}}
{ {\bar{h}^{\lambda}} }_{(\bar{N})\mu_{0} \rho_{0}} \right) S^{\rho}_{\nu} \quad
\end{equation}
and also
\begin{equation} \label{KN2}
S^{\mu}_{\sigma}K^{(\epsilon_{P}) \,\lambda}_{(N^2)\sigma \nu} = -\epsilon_{P} i \left(h^{\lambda}_{(N)\mu_{1} \rho_{1}}\delta^{\mu_{0}}_{ \rho_{0}} - \delta^{\mu_{1}}_{\rho_{1}}{ {\bar{h}^{\lambda}} }_{(\bar{N})\mu_{0}\rho_{0}} \right) S^{\rho}_{\nu} \quad,
\end{equation}
where the single index $\mu$ appearing on the left corresponds to the double indices $\mu_{1}\mu_{0}$ appearing on the right. A convenient correspondence is given by
\begin{equation} \label{mumu1mu0}
\mu = N(\mu_{1}-1)+\mu_{0} \quad ; \quad \rho = N(\rho_{1}-1)+\rho_{0} \quad,
\end{equation}
with $\mu \in$ $\{1,...,N^2\}$ and $\mu_{1},\mu_{0} \in$ $\{1,...,N\}.$ Likewise for $\rho.$ A proof of (\ref{JN2}) and (\ref{KN2}) can be found in the Appendix. 

We turn back now to the general case. To continue, we work in the $(A,B)\oplus (C,D)$ rep of the $N$-Lorentz-Weyl algebra and define the following matrices 
 \begin{equation} \label{ABCD1} J^{\mu}_{(ABCD)} \equiv \pmatrix{J^{\mu}_{(A,B)} && 0 \cr 0 && J^{\mu}_{(C,D)}}  \quad ; \quad K^{\mu}_{(ABCD)} \equiv \pmatrix{K^{\mu}_{(A,B)} && 0 \cr 0 && K^{\mu}_{(C,D)}} \quad,
 \end{equation}
 \begin{equation} \label{ABCD2} P^{(\epsilon_{P})\,\mu}_{(ABCD)} = \pmatrix{0 && P_{+}^{(\epsilon_{P})\,\mu} \cr P_{-}^{(\epsilon_{P})\,\mu} && 0}  \quad .
 \end{equation}
Compare (\ref{ABCD1}) and (\ref{ABCD2}) with (\ref{anti1}) and (\ref{anti2}). Both here and there, the $J$s and $K$s are block diagonal while the $P$s have off-diagonal blocks. By the off-diagonal structure of definition (\ref{ABCD2}), translations generated with ${P}^{\mu}_{+}$ change quantities in $(A,B)$-space and leave quantities transforming with $(C,D)$ invariant. And ${P}^{\mu}_{-}$ changes $(C,D)$ quantities and leaves $(A,B)$ quantities invariant.

We can now find matrices $P^{(\epsilon_{P})\,\mu}_{(ABCD)}$ that, with the known matrices $J^{\mu}_{(ABCD)}$ and $K^{\mu}_{(ABCD)},$ obey the $N$-Poincar\'{e}-Weyl commutation relations (\ref{Pcomm1}). By a straightforward calculation, we combine the fact that the matrices $J_{(N^2)}$ are the coefficients $if$ in (\ref{JKN2}) with the fact that $J_{(N^2)}$ and $K_{(N^2)}$ are equivalent to the $(N,\bar{N})$ rep by a similarity transformation $S$ and put these facts in the $[P,J]$ and $[P,K]$ $N$-Poincar\'{e}-Weyl commutation relations (\ref{Pcomm1}). 

Putting all this together, one finds equations that $P^{(\epsilon_{P})\,\mu}_{(ABCD)}$ must obey. The $P^{(\epsilon_{P})\,\mu}_{(ABCD)}$s are in block form (\ref{ABCD2}) with nonzero blocks $P_{+}$ and  $P_{-}.$ One finds $P_{+}$ block equations from the $[P,J]$ = $ifP$ commutation relation,
$$ \tilde{P}^{(\epsilon_{P})\,\lambda}_{+\,abst}\left(J^{\nu}_{(C)\,sc}\delta^{t}_{d} + \delta^{s}_{c}J^{\nu}_{(D)\,td}\right) = \hspace{10cm}$$
$$ \hspace{1cm} =\left[ \left( h^{\nu}_{(N)\,\lambda_{1} \rho_{1}}\delta^{\lambda_{0}}_{\rho_{0}} + \delta^{\lambda_{1}}_{\rho_{1}}\bar{h}^{\nu}_{(\bar{N})\,\lambda_{0} \rho_{0}}\right)\delta^{rs}_{ab} +\left( J^{\nu}_{(A)\,ar} \delta^{s}_{b} +\delta^{r}_{a} J^{\nu}_{(B)\,bs} \right) \delta^{\lambda}_{\rho} \right]\tilde{P}^{(\epsilon_{P})\,\rho_{1}\rho_{0}}_{+ \,rscd} \quad ,$$
where single and double indices are related by (\ref{mumu1mu0}) and we define $\tilde{P}$ by
\begin{equation} \label{SimP} \tilde{P}^{\lambda_{1}\lambda_{0}} = \tilde{P}^{\lambda} \equiv S^{\lambda}_{\mu}P^{\mu} \quad,
\end{equation}
where  $S^{\lambda}_{\mu}$ is the similarity transformation in (\ref{JN2}) and (\ref{KN2}).
From the $[P,K]$ = $-\epsilon_{P}idP$ commutation relation, we get
$$ \tilde{P}^{(\epsilon_{P})\,\lambda}_{+\,abst}\left(J^{\nu}_{(C)\,sc}\delta^{t}_{d} - \delta^{s}_{c}J^{\nu}_{(D)\,td}\right) = \hspace{10cm}$$
$$ \hspace{1cm} =\left[ \epsilon_{P}\left( h^{\nu}_{(N)\,\lambda_{1} \rho_{1}}\delta^{\lambda_{0}}_{\rho_{0}} - \delta^{\lambda_{1}}_{\rho_{1}}\bar{h}^{\nu}_{(\bar{N})\,\lambda_{0} \rho_{0}}\right)\delta^{rs}_{ab} +\left( J^{\nu}_{(A)\,ar} \delta^{s}_{b} - \delta^{r}_{a} J^{\nu}_{(B)\,bs} \right)\delta^{\lambda}_{\rho} \right]\tilde{P}^{(\epsilon_{P})\,\rho_{1}\rho_{0}}_{+ \,rscd} \quad .$$
By adding and subtracting the above two equations we get
\begin{equation} \label{ABCD3} \tilde{P}^{(\epsilon_{P})\,\lambda}_{+\,absd}J^{\nu}_{(C)\,sc}  = \left[\left(\frac{1}{2}(1+\epsilon_{P})h^{\nu}_{(N)\,\lambda_{1} \rho_{1}}\delta^{\lambda_{0}}_{\rho_{0}} + \frac{1}{2}(1-\epsilon_{P})\delta^{\lambda_{1}}_{\rho_{1}}\bar{h}^{\nu}_{(\bar{N})\,\lambda_{0} \rho_{0}}\right)\delta^{rs}_{ab} + J^{\nu}_{(A)\,ar}\delta^{s}_{b}\delta^{\lambda}_{\rho}\right]\tilde{P}^{(\epsilon_{P})\,\rho_{1}\rho_{0}}_{+ \,rscd}
\end{equation}
$$\tilde{P}^{(\epsilon_{P})\,\lambda}_{+\,abct}J^{\nu}_{(D)\,td}  = \left[\left(\frac{1}{2}(1-\epsilon_{P})h^{\nu}_{(N)\,\lambda_{1} \rho_{1}}\delta^{\lambda_{0}}_{\rho_{0}} + \frac{1}{2}(1+\epsilon_{P})\delta^{\lambda_{1}}_{\rho_{1}}\bar{h}^{\nu}_{(\bar{N})\,\lambda_{0} \rho_{0}}\right)\delta^{rs}_{ab} + \delta^{r}_{a} J^{\nu}_{(B)\,bs}\delta^{\lambda}_{\rho}\right]\tilde{P}^{(\epsilon_{P})\,\rho_{1}\rho_{0}}_{+ \,rscd}$$
Thus, for $\epsilon_{P}$ = +1, the $C$-rep is related to the fundamental rep ($N$) and the $A$-rep, while $D$ is related to the anti-rep ($\bar{N}$) and $B.$ 

In terms of rotations $D(\theta)$ = $\exp{(i\theta_{\mu}J^{\mu})},$ and with $\epsilon_{P}$ = $+1,$ (\ref{ABCD3}) gives
\begin{equation} \label{ABCD4} \tilde{P}^{(\epsilon_{P})\,\lambda_{1}\lambda_{0}}_{+\,absd} D^{(C)}_{sc}(\theta) = D^{(N)}_{\lambda_{1} \rho_{1}}(\theta)D^{(A)}_{ar}(\theta) \tilde{P}^{(\epsilon_{P})\,\rho_{1}\lambda_{0}}_{+\,rbcd}  \quad \quad [\epsilon_{P} = +1] \quad
\end{equation}
$$  \tilde{P}^{(\epsilon_{P})\,\lambda_{1}\lambda_{0}}_{+\,abcs} D^{(D)}_{sd}(\theta) = D^{(\bar{N})}_{\lambda_{0} \rho_{0}} (\theta)D^{( B)}_{bs} (\theta)\tilde{P}^{(\epsilon_{P})\,\lambda_{1}\rho_{0}}_{+\,ascd}   \quad . \quad [\epsilon_{P} = +1] \quad  $$
By definition, the components of $\tilde{P}_{+}$ are the Clebsch-Gordan coefficients relating the $C$-rep of U($N$) with the $N \otimes A$ direct product rep and the $D$-rep with the $\bar{N} \otimes B$ rep.\cite{Hamermesh} Therefore the components of $\tilde{P}_{+}$ are products of Clebsch-Gordan coefficients,
\begin{equation} \label{ABCD5} \tilde{P}^{(\epsilon_{P})\,\lambda}_{+\,abcd}  = \tilde{P}^{(\epsilon_{P})\,\lambda_{1}\lambda_{0}}_{+\,abcd} = k_{+} (N\lambda_{1}Aa \mid Cc)(\bar{N}\lambda_{0}Bb \mid Dd) \quad ,  \quad [\epsilon_{P} = +1] \quad
 \end{equation}
where $k_{+}$ is a constant and we use (\ref{mumu1mu0}) to relate the $\lambda$s, i.e. $\lambda$ = $N(\lambda_{1}-1)+\lambda_{0}.$  

For the $\tilde{P}_{-}$ block, we find similarly that 
\begin{equation} \label{ABCD6} \tilde{P}^{(\epsilon_{P})\,\lambda}_{-\,cdab}  = \tilde{P}^{(\epsilon_{P})\,\lambda_{1}\lambda_{0}}_{-\,cdab} = k_{-} (N\lambda_{1}Cc \mid Aa)(\bar{N}\lambda_{0}Dd \mid Bb) \quad .  \quad [\epsilon_{P} = +1] \quad
 \end{equation}
Thus the matrices $\tilde{P}^{\mu}$ are arrays of Clebsch-Gordan coefficients and the momentum matrices ${P}^{\mu}$ can be obtained by the inverse of the similarity transformation $S,$  
\begin{equation} \label{PSP} P^{\mu} = {S^{-1}}^{\mu}_{\sigma}\tilde{P}^{\sigma} \quad .
 \end{equation}
For $\epsilon_{P}$ = $-1$ exchange basic reps $N$ and $\bar{N}$ in the above discussion.

In order for the momentum matrices to commute, $[{P}^{(\epsilon_{P})\,\mu}_{(ABCD)},{P}^{(\epsilon_{P})\,\nu}_{(ABCD)}]$ = 0, we make one of the off-diagonal blocks vanish,
 \begin{equation} \label{ABCD6a}   {P}^{(\epsilon_{P})\,\mu}_{+\,abcd}  = 0  \quad {\mathrm{or}} \quad {P}^{(\epsilon_{P})\,\mu}_{-\,cdab}  = 0 \quad ,
 \end{equation}
 i.e. $k_{+} $ = 0 or $k_{-} $ = 0. This completes the process of finding the momentum matrices $P^{(\epsilon_{P})\,\mu}_{(ABCD)}$ that satisfy the $N$-Poincar\'{e}-Weyl algebra (\ref{Pcomm1}) with the $J$s and $K$s of the $(A,B)\oplus(C,D)$ rep of the $N$-Lorentz-Weyl algebra (\ref{Lorentz}).

When applied to the $(N^2)$ rep (\ref{JKN2}) for $N^2$-dimensional spacetime, we have $(A,B)$ = $(N,\bar{N})$ for $\epsilon_{P}$ = $+1,$ as discussed above with (\ref{JN2}) and (\ref{KN2}). To get momentum matrices that produce nontrivial, faithful translations in $N^2$-dimensional spacetime, we must combine the $(N^2)$ rep with another rep $(C,D),$  $(A,B)\oplus (C,D)$ = $(N,\bar{N})\oplus (C,D).$ As just shown, we get momentum matrices when there are nonzero Clebsch-Gordan coefficients. For the $P_{+}$ block this means when 
 \begin{equation} \label{CNN+} C \in N \otimes A = N \otimes N \quad {\mathrm{and}} \quad D \in \bar{N} \otimes B = \bar{N} \otimes \bar{N} \quad ( \epsilon_{P} = +1) \quad.
 \end{equation}
And for the $P_{-}$ block we get nonzero Clebsch-Gordan coefficients when 
 \begin{equation} \label{CNN-} A = N \in N \otimes C \quad {\mathrm{and}} \quad B = \bar{N} \in \bar{N} \otimes D \quad (P^{-}\; {\mathrm{and}} \; \epsilon_{P} = +1) \quad.
 \end{equation}
[By `$ H \in G$' we mean the rep $H$ is included in the sum of irreducible reps that make up the rep $G.$]

Thus, $N^2$-dimensional spacetime on its own does not have any momentum matrices, no irreducible representation of the $N$-Lorentz-Weyl algebra does. To have momentum matrices and therefore translation matrices, $N^2$-dimensional spacetime must be combined with a second manifold whose transformations form a suitable representation of the $N$-Lorentz-Weyl algebra.

\section{The Utility Rep; N-squared Dimensional Spacetime} \label{UNspacetime}

The work so far has been very general. In this section a particular basis is described that has tools for understanding the characteristics of the transformations found in  spacetime. Call it the `utility rep.' This representation is well known.\cite{specialREP}  In the utility rep, time and space are clearly distinguished, there is a way to consider the basis matrices $h^{\mu}_{(N)}$ to be orthonormal, the coefficients $f^{\mu\lambda \nu}$ are completely antisymmetric in $\mu\lambda \nu,$ the coefficients $d^{\mu\lambda \nu}$ are completely symmetric and scale transformations appear as a kind of boost. 

The $J$ and $K$ generators, in the utility rep, are denoted by lower case letters: $J \rightarrow j$ and $K \rightarrow k.$ Translations require reducible reps of the $N$-Lorentz-Weyl commutation relations and the utility rep used in this section is an irreducible rep. Therefore translations and their generators $P$ are not considered in this section. 

Let $C^{ab}$ be the $N\times N$ matrix with zero for all components except the $ab$th component which is one. We write the $mn$th component as $C^{ab}_{mn}$ = $\delta^{ab}_{mn},$ where $\delta$ is one when the upper sequence of indices $ab...$ is equal to the lower sequence of indices $mn...$ and zero otherwise. 

 Define the matrices $h^{\epsilon ab}_{(N)}$ that form a basis of the $N \times N$ hermitian matrices by
  $$   h^{+, ab}_{(N)}  = \frac{1}{2}\left( C^{ab} + C^{ba} \right)  \quad  {\mathrm{;}} \quad h^{-, ab}_{(N)}  = \frac{-i}{2}\left( C^{ab} - C^{ba} \right) \quad 1 \leq a<b \leq N $$ 
\begin{equation} \label{AJN} h^{\,0, 11}_{(N)}  = \frac{1}{\sqrt{2N}}\;{\mathbf{1}} \quad {\mathrm{;}} \quad h^{\,0, aa}_{(N)}  = \frac{1}{\sqrt{2(a^2-a)}}\left( \sum^{a-1}_{n=1}C^{nn} -(a-1) C^{aa} \right)  \quad a \in \{2,...,N\}    \quad ,
 \end{equation}
where the boldface ${\mathbf{1}}$ indicates the $N\times N$ identity matrix, ${\mathbf{1}}_{mn}$ = $\delta^{m}_{n}.$ 

There are $N(N-1)/2$ + $N(N-1)/2$ + $1$ + $N-1$  = $N^2$ matrices $h^{\epsilon, ab}_{(N)}.$ The $h^{\epsilon, ab}_{(N)}$ are hermitian and all but $h^{0,11}_{(N)}$ are traceless. The label `$\epsilon, ab$' is awkward, so let us relabel with the position $\sigma$ of $h^{\epsilon, ab}_{(N)}$ in the following list,  
$$h^{\sigma}_{(N)} \in \{h^{+, ab}_{(N)},\; h^{-, ab}_{(N)},\; h^{0, aa}_{(N)},\; h^{0,11}_{(N)}\} \quad {\mathrm{and}} \quad \bar{h}^{\sigma}_{(N)} \in \{-h^{+, ab}_{(N)},\; +h^{-, ab}_{(N)},\; -h^{0, aa}_{(N)},\; -h^{0,11}_{(N)}\} \quad ,$$ 
where the index $\sigma \in$ $\{1,...,N^2\}$ and the ordering is the same for both bases $h^{\sigma}_{(N)}$ and $\bar{h}^{\sigma}_{(N)}.$  

Space and time are distinguished by the trace of the basis matrices. The spatial $h^{\mu}_{(N)}$s are the $N^2 - 1$ traceless matrices $h^{i}_{(N)}.$ The matrix $h^{N^2}_{(N)}$ = $h^{t}_{(N)}$ is the time component and is the only matrix of the $h^{\mu}_{(N)}$s that has a non-zero trace.
$$ {\mathrm{tr}}(h^{i}_{(N)}) = 0 \quad {\mathrm{and}} \quad {\mathrm{tr}}(h^{t}_{(N)}) = {\mathrm{tr}}(h^{N^2}_{(N)}) = \sqrt{N/2} \quad , $$
where the spatial indices are $i \in$ $\{1,...,N^2-1\}$ and $\mu$ = $N^2$ = $t$ is the time index.

One can show that the $h^{\sigma}_{(N)}$s obey the trace identity,
\begin{equation} \label{trace1} 2{\mathrm{tr}}(h^{\mu}_{(N)}h^{\nu}_{(N)}) = \delta^{\mu}_{\nu} \quad ,
\end{equation}
This orthonormality gives a way to find coefficients in a sum,
\begin{equation} \label{ortho} A = \sum \alpha_{\sigma} h^{\sigma}_{(N)} \quad {\mathrm{implies}} \quad  \alpha_{\mu} = 2 {\mathrm{tr}}(A h^{\mu}_{(N)}) \quad,
 \end{equation}
where matrix multiplication in $A h^{\mu}_{(N)}$ is implied by the context, just as it is elsewhere in this article. 

We can use the orthonormality (\ref{ortho}) to determine the coefficients in the fundamental commutators and anticommutators (\ref{Comm1}), 
\begin{equation} \label{fdTRACE} f^{\mu \lambda \nu} = -2i{\mathrm{tr}}([h^{\mu}_{(N)},h^{\lambda}_{(N)}]h^{\nu}_{(N)}) \quad {\mathrm{and}} \quad d^{\mu \lambda \nu} = 2{\mathrm{tr}}(\{h^{\mu}_{(N)},h^{\lambda}_{(N)}\}h^{\nu}_{(N)}) \quad .
\end{equation}
Since the trace of a matrix product $AB$ is independent of the order, ${\mathrm{tr}}(AB)$ = $A_{ms}B_{sm}$ = $B_{sm}A_{ms}$ =  ${\mathrm{tr}}(BA),$ one can show that (\ref{fdTRACE}) implies
\begin{equation} \label{fAntidSymm} f^{\mu \lambda \nu} = -f^{\lambda \mu \nu} = -f^{\mu \nu \lambda} \quad {\mathrm{and}} \quad d^{\mu \lambda \nu} = +d^{\lambda \mu \nu} = +d^{\mu \nu \lambda} \quad ,
\end{equation}
which shows that $f^{\mu \lambda \nu}$ is antisymmetric under exchange of any two of the indices $\mu \lambda \nu$ and $d^{\mu \lambda \nu}$ is symmetric under exchange of any two of the indices $\mu \lambda \nu.$ Thus $f^{\mu \lambda \nu}$ is completely antisymmetric and $d^{\mu \lambda \nu}$ is completely symmetric in this special rep.

The generators of rotations and boosts in $N^2$-dimensional spacetime are just the coefficients $if$ and $id,$ see (\ref{JKN2}),
\begin{equation} \label{jKN2} 
{j^{\lambda}_{(N^2)\, {\mu \nu}}}  =  i f^{\mu \lambda \nu}  \quad ; \quad {k^{(\epsilon_{P}) \, \lambda}_{(N^2)\, \mu\nu}}  =  -\epsilon_{P} i d^{\mu \lambda \nu} \quad.
\end{equation}
By the symmetries of $f$ and $d$ in (\ref{fAntidSymm}), the angular momentum matrices $j^{\lambda}_{(N^2)}$ are antisymmetric in $\mu \nu$ and the boost matrices $k^{\lambda}_{(N^2)}$ are symmetric in $\mu \nu$. Thus the $j^{\lambda}_{(N^2)}$s are hermitian and the $k^{\lambda}_{(N^2)}$s are anti-hermitian. Furthermore, rotations are unitary and boosts are not.

The time index angular momentum matrix $j^{t}_{(N^2)}$ and the time index boost matrix $k^{t}_{(N^2)}$ are especially simple. Since the time index basis matrix, $h^{t}_{(N)} \equiv$   $h^{N^2}_{(N)}$ = ${\mathbf{1}}/\sqrt{2N},$ is proportional to the unit matrix, its commutators and anticommutators are simple. We find that
$$ [h^{\mu}_{(N)},h^{t}_{(N)}] = \frac{1}{\sqrt{2N}}[h^{\mu}_{(N)},{\mathbf{1}}] = {\mathbf{0}} \quad {\mathrm{and}} \quad \{h^{\mu}_{(N)},h^{t}_{(N)}\} = \frac{1}{\sqrt{2N}} \{h^{\mu}_{(N)},{\mathbf{1}}\} =  \sqrt{\frac{2}{N}} h^{\mu}_{(N)} \quad ,$$ 
which implies, by (\ref{fdTRACE}) and (\ref{jKN2}), that 
\begin{equation} \label{jKN2N2} 
{j^{t}_{(N^2)\, {\mu \nu}}}  =  i f^{\mu t \nu} =  0 \quad {\mathrm{and}} \quad {k^{(\epsilon_{P}) \, t}_{(N^2)\, \mu\nu}} =  -\epsilon_{P} i d^{\mu t \nu}  =  -\epsilon_{P} i \sqrt{\frac{2}{N}} \;  \delta^{\mu}_{\nu} = -\epsilon_{P} i \sqrt{\frac{2}{N}} \; {\mathbf{1}}_{\mu \nu}\quad,
\end{equation}
where $ {\mathbf{1}}$ is now the $N^2 \times N^2$ unit matrix. Thus the time index angular momentum matrix vanishes, while the time index boost matrix is proportional to the unit matrix.
 
Now consider a rotation followed by a boost.  Let 
\begin{equation} \label{Dtrans}
D^{\epsilon_{P}}_{(N^2)}(\theta,\phi) \equiv \exp{(i\phi_{\sigma}k^{(\epsilon_{P}) \, \sigma}_{(N^2)})}\exp{(i\theta_{\rho}j^{\rho}_{(N^2)})}
\end{equation}
 be the $N^2$-dimensional $N$-Lorentz-Weyl transformation matrix for a rotation through angle $\theta$ = $\{\theta_{1},...,\theta_{N^2}\}$ followed by a boost through `boost parameter' $\phi$ = $\{\phi_{1},...\phi_{N^2}\}.$ 

We can introduce $N^2$-dimensional spacetime coordinates as all ordered sets of real numbers $x^{\mu},$ $x^{\mu} \in$ $\{x^{1},...,x^{N^2}\}.$ In the special rep (\ref{AJN}) described above, a rotation followed by a boost associates new coordinates with the original coordinates,
\begin{equation} \label{xprime}
{x^{\prime}}^{\mu} = \left[D^{\epsilon_{P}}_{(N^2)}(\theta,\phi) \right]_{\mu \nu} x^{\nu}
\end{equation}
Using the language of spacetime here, one can say that the coordinates $x^{\mu}$ locate an `event' and the ${x^{\prime}}^{\mu}$ are coordinates of the same event after rotating and boosting.

From the antisymmetry (\ref{jKN2}) in the angular momentum matrices, $j^{\lambda}_{(N^2)\; \mu \nu}$ = $-j^{\lambda}_{(N^2)\; \nu \mu},$ it follows that the squares of `distances' such as $\sum_{i}{x^{i}}^2$ and $\sum_{i}{(x^{i}-y^{i})}^2$ are invariant under rotations. To show this, consider an infinitesimal angle $\delta \theta_{\mu}$ and null $\phi_{\sigma}$ = 0 in (\ref{Dtrans}), we have
\begin{equation} \label{xprime2}
\sum_{i}{{x^{\prime}}^{i}}^2 =  (\delta_{i \nu} + i\delta \theta_{\rho} j^{\rho}_{(N^2) \; i \nu})x^{\nu}(\delta_{i \bar{\nu}} + i\delta \theta_{\rho} j^{\rho}_{(N^2) \; i \bar{\nu}})x^{\bar{\nu}} \quad {\mathrm{(prime:}} \; {\mathrm{ rotation)}}
\end{equation}
$$ = \delta_{i \nu}\delta_{i \bar{\nu}}x^{\nu}x^{\bar{\nu}} + 2i\delta \theta_{\rho} j^{\rho}_{(N^2) \; i j}x^{i}x^{{j}} =  \sum_{i}{x^{i}}^2 \quad ,
$$
where we display the sum signs because the index `$i$' is not repeated in the square ${x^{i}}^2.$ The combination $j^{\rho}_{(N^2) \; i j}x^{i}x^{{j}}$ vanishes because  the $j^{\rho}_{(N^2) \; i j}$s are antisymmetric while the $x^{i}x^{j}$s are symmetric in $ij.$ 

Rotations preserve the time components of vectors. Note that the complete antisymmetry of $if^{\mu \lambda \nu}$ together with $if^{\mu \, t \, \nu}$ = 0, implies $if^{ t \, \mu \nu}$ = 0 and $if^{ \mu \nu \, t}$ = 0. Since the $if$s are the components of angular momentum matrices, we have both $j^{\mu}_{(N^2)\; t \, \nu}$ = 0 and $j^{\lambda}_{(N^2) \; \nu \, t}$ = 0. Thus the time components of the angular momentum matrices vanish and rotations do not change time. We have
\begin{equation} \label{tprime}
{x^{\prime}}^{t} =   x^{t}\quad , \quad  {\mathrm{(prime:}} \; {\mathrm{ rotation)}} \quad ,
\end{equation}
where we use the label `$t$' to emphasize the time $x^{t} \equiv$ $x^{N^2}.$ Thus we have shown that both time coordinates as well as distances are invariant under rotations.

For $N \neq$ 2, i.e. $N \geq$ 3, boosts do not preserve general spacetime intervals, which when squared is the distance squared minus time squared, $\sum_{i}{x^{i}}^2 - {x^{t}}^2.$ In fact, with an infinitesimal boost, $\delta \phi_{\mu},$  one finds that
 \begin{equation} \label{xt2}
\sum_{i}{{x^{\prime}}^{i}}^2 - {{x^{\prime}}^{t}}^2 =  \sum_{i}{x^{i}}^{2} -{x^{t}}^{2} + 2 \epsilon_{P}\delta \phi_{\rho}( d^{i \rho \, j}x^{{i}}x^{j} -  d^{t \rho \, t}{x^{t}}^2) \quad {\mathrm{(prime:}} \; {\mathrm{ boost)}}
\end{equation}
By examining the values of $d^{i \rho \, j}$ and $d^{t \rho \, t}$ for various $N,$ one conjectures that the spacetime interval is not invariant under boosts for $N \geq$ 3. 

By selecting boost parameters $\delta\phi_{\rho}$ and selecting corresponding spacetime subspaces, one can arrange for the offending quantity $\delta \phi_{\rho}(\sum_{i,j} d^{i \rho \, j}x^{{i}}x^{j} -  d^{t \rho \, t}{x^{t}}^2)$ to vanish and in such subspaces for boosts in the selected directions, the spacetime intervals are invariant. 

For example, consider the basis $h_{(N)}$s indexed as they were defined in (\ref{AJN}). The three matrices $h^{+, 12}_{(N)}$ $h^{-, 12}_{(N)}$ and $h^{0, 22}_{(N)}$ are $2 \times 2$ Pauli spin matrices each with zeros added to fill out an $N \times N$ matrix. It follows that for the $3+1$ spacetime $\{x,y,z,t\}$ = $\{x^{+, 12},x^{+, 12},x^{0, 22},x^{0, 11}\}$ to have invariant spacetime intervals, $\sum{{x^{i}}^2 - {x^{t}}^2},$ it is sufficient that nonzero boost parameters $\delta \phi$ are restricted to $\{\delta \phi_{x},\delta \phi_{y},\delta \phi_{z},0,...,0 \}.$

For $N$ = 2, there are just four components, so to have invariant spacetime intervals we need $\{\delta \phi_{x},\delta \phi_{y},\delta \phi_{z},0 \}$ i.e. we just need to specify $\delta \phi_{t}$ = 0 so that there are no scale transformations. This case is four dimensional spacetime and all spacetime intervals are invariant under all space-directed boosts. The case is discussed in the next section. 

The $N^2$-dimensional spacetime transformations include a boost in the time direction, the boost generated by the matrix $k^{(\epsilon_{P}) \, t}_{(N^2)}$ = $-\epsilon_{P} i \sqrt{2/N} \; {\mathbf{1}}$ in (\ref{jKN2N2}). Thus with all boost parameters zero except for $\phi_{t},$ the time-directed boost is multiplication by a scale factor,
 \begin{equation} \label{tboost}
{x^{\prime}}^{\mu}  = \exp{(i\phi_{t}k^{(\epsilon_{P})\, t})} =   e^{(\epsilon_{P} \phi_{t} \sqrt{2/N})} x^{\mu} \quad , \quad {\mathrm{(prime:}} \; {\mathrm{ time \; boost)}}
\end{equation}
where $\theta$ = 0 and $\phi$ = $\{0,0,...,\phi_{t}\}.$  Scale transformations were studied by Weyl,\cite{Weyl} which motivates appending his name to the Lorentz and Poincar\'{e} commutation relations.

\section{4-d Spacetime; $N$ = 2} \label{N=2}

We continue with the utility rep discussed in the last section, specialized further to the case $N$ = 2 and $\epsilon_{P}$ = +1; the discussion for $\epsilon_{P}$ = $-1$ is similar. One finds the rotation, boosts, and scale transformations of four-dimensional spacetime.

The basis of hermitian $2\times 2$ matrices from (\ref{AJN}) are the following matrices
 \begin{equation} \label{j4N=2}
h^{1} = h^{x} = h^{+12} =  \frac{1}{2}\pmatrix{0 && 1 \cr 1 && 0} \quad ; \quad h^{2} = h^{y} = h^{-12} =  \frac{1}{2}\pmatrix{0 && -i \cr i && 0} \quad ; 
\end{equation}
$$ h^{4} = h^{t} = h^{011} =  \frac{1}{2}\pmatrix{1 && 0 \cr 0 && 1}  \quad ; \quad h^{3} = h^{z} = h^{022} =  \frac{1}{2}\pmatrix{1 && 0 \cr 0 && -1} \quad . 
$$
In this section, we sometimes use the indices $x,y,z,t$ and sometimes we use the numerical $1,2,3,4;$ they are interchangeable. The notation $\epsilon ab$ index refers back to the original definition of the utility basis in (\ref{AJN}). Direct calculations show these matrices satisfy the orthonormality condition (\ref{ortho}), $2{\mathrm{tr}}(h^{\mu}h^{\nu})$ = $\delta^{\mu}_{\nu},$  for $\mu,\nu \in$ $\{1,2,3,4\}.$ 

By (\ref{j4N=2}) the coefficients $f^{\mu \nu \sigma}$ in (\ref{Comm1}), i.e. $[h^{\mu},h^{\nu}]$  = $i f^{\mu \nu \sigma} h^{\sigma},$ are found to be
 \begin{equation} \label{f4N=2}
f^{\mu \lambda \nu} = \epsilon^{\mu \lambda \nu}\quad , 
\end{equation}
where $\epsilon^{\mu \lambda \nu }$ is the completely antisymmetric symbol with $\epsilon^{xyz}$ = 1 and $\epsilon^{\mu \lambda \nu}$ vanishes for any $\mu,\lambda,\nu$ = $t$, so $\epsilon^{t \lambda \nu}$ = $\epsilon^{\mu t \nu}$ = $\epsilon^{\mu \lambda t}$ = 0. 
The coefficients $d^{\mu \nu \sigma}$ in $ \{h^{\mu},h^{\nu}\}$  =  $d^{\mu \nu \sigma} h^{\sigma}$ are found by direct calculation from (\ref{j4N=2}) to be
 \begin{equation} \label{d4N=2}
d^{\mu \lambda \nu} =  \delta^{\mu t}_{\lambda \nu} +\delta^{\lambda t}_{\nu \mu} + \delta^{\nu t}_{\mu \lambda} - 2 \delta^{\mu \lambda \nu}_{t\,t\,t} \quad . 
\end{equation}
Clearly, the $f^{\mu \lambda \nu}$ are completely antisymmetric in $\mu, \lambda, \nu$ and the $d^{\mu \lambda \nu}$ are completely symmetric in $\mu, \lambda, \nu.$ 

By (\ref{JKN2}) and (\ref{jKN2}), the angular momentum matrices are proportional to the $f$s,
\begin{equation} \label{j4N=2a} 
{j^{i}_{(4)\, {\mu \nu}}}  =  i f^{\mu i \nu} = i\epsilon^{\mu i \nu} \quad ; \quad {j^{t}_{(4)\, {\mu \nu}}}  = 0 \quad,
\end{equation}
where $i \in$ $\{x,y,z\}$ and $\mu, \nu \in$ $\{x,y,z,t\}.$ The angular momentum matrices, ${j^{x}_{(4)}},{j^{y}_{(4)}},{j^{z}_{(4)}},$ generate the usual spacial rotations.  For example, consider the rotation of the $xy$-plane through an angle $\theta$ = $\{0,0,\theta_{z},0\}.$ The rotation matrix is generated by ${j^{z}_{(4)}};$ one finds by (\ref{Dtrans}) and (\ref{j4N=2a}),
\begin{equation} \label{Rot4N=2}
D_{(4)}(\theta,0) = \exp{(i\theta_{z} j^{z}_{(4)})} = \pmatrix{\cos{\theta_{z}} && \sin{\theta_{z}} &&0 &&0 \cr -\sin{\theta_{z}} && \cos{\theta_{z}} &&0 &&0 \cr   0 && 0 &&1 &&0 \cr 0 && 0 &&0 &&1 } \quad. \quad [\theta = \{0,0,\theta_{z},0\}]
\end{equation}
One recognizes the rotation matrix for a rotation in the $xy$-plane with invariant $z$ and $t.$

The boost matrices are proportional to the $d$s, by (\ref{JKN2}). We get
\begin{equation} \label{j4N=2b} 
{k^{ i}_{(4)\, {\mu \nu}}}  =  -\epsilon_{P} i d^{\mu i \nu} = -i\left(\delta^{i t}_{\mu \nu} +\delta^{i t}_{\nu \mu} \right)  \quad ; \quad {k^{ t}_{(4)\, {\mu \nu}}}  =  -i\delta^{\nu}_{\mu } =   -i{\mathbf{1}}_{\mu \nu} \quad,
\end{equation}
where we drop the label $(\epsilon_{P})$ in $k^{(\epsilon_{P}) \, i}_{(4)}$ because we take $\epsilon_{P}$ = +1 in this section. The quantity ${\mathbf{1}}$ is the $4 \times 4$ unit matrix. The spatial-directed boost matrices, ${k^{ x}_{(4)}},{k^{ y}_{(4)}},{k^{ z}_{(4)}},$ generate the usual boosts.  For example, consider the boost in the $z$-direction with boost parameter $\phi$ = $\{0,0,\phi_{z},0\},$ one finds by (\ref{Dtrans}) and (\ref{j4N=2b}),
\begin{equation} \label{Boost4N=2}
D_{(4)}(0,\phi) = \exp{(i\phi_{z}k^{ z}_{(4)})} = \pmatrix{1 && 0 &&0 &&0 \cr 0 && 1 &&0 &&0 \cr   0 && 0 && \cosh{\phi_{z}} && \sinh{\phi_{z}} \cr 0 && 0 &&\sinh{\phi_{z}} &&\cosh{\phi_{z} }} \; . \; [\phi = \{0,0,\phi_{z},0\}]
\end{equation}
Thus we recover the well-known rotations and boosts of 4-d spacetime.

Now consider an infinitesimal boost without a scale transformation, so that we have $\phi$ = $\{\delta \phi_{x},\delta \phi_{y},\delta \phi_{z},0\}.$ We put $\delta \phi_{\, t}$ = 0, so there are no scale transformations. Then the infinitesimal change in the square of the spacetime interval is given by (\ref{xt2}),
 $$
\sum_{i}{{x^{\prime}}^{i}}^2 - {{x^{\prime}}^{t}}^2 =  \sum_{i}{x^{i}}^{2} -{x^{t}}^{2} +  2 \delta \phi_{j}(\sum_{i} d^{i j \, k}x^{{i}}x^{k} -  d^{t j \, t}x^{{t}}x^{t}) =  \sum_{i}{x^{i}}^{2} -{x^{t}}^{2} \quad . \quad {\mathrm{(prime:}} \; {\mathrm{ boost)}}
$$
Note that, by (\ref{d4N=2}), the  coefficients $d^{ijk}$ vanish since $i,j,k$ are all spatial indices and the coefficients $d^{t j t}$ vanish because two of the indices are timelike while the third is not. Thus, for $N$ = 2 and not allowing scale transformations, 4-d spacetime has spacetime intervals preserved under rotations and boosts.

Since ${j^{t}_{(4)}}$ vanishes, the rotation, $\exp{(i \theta_{t} j^{t}_{(4)})},$ is just the unit matrix which leaves any 4-vector unchanged. Thus the `rotations' generated by $j^{t}_{(4)}$ are trivial, multiplication by unity.

The time-index boost matrix ${k^{ t}_{(4)}}$ is proportional to the unit matrix, ${k^{ t}_{(4)}}$ = $-i{\mathbf{1}}$ generating a boost in the time direction that is proportional to the unit matrix. For $\phi$ = $\{0,0,0,\phi_{t}\},$ we find by (\ref{Dtrans}) that
\begin{equation} \label{timeBoost4N=2}
D_{(4)}(0,\phi) = \exp{(i\phi_{t}k^{ t}_{(4)})} = \exp{(i\phi_{t}(-i{\mathbf{1}}))} = e^{\phi_{t}} {\mathbf{1}} \; . \quad [\phi = \{0,0,0,\phi_{t}\}]
\end{equation}
Thus the time-directed boost multiplies every vector component by the same factor, $\exp{\phi_{t}}  $, and so the time-directed boost is a scale transformation. 

In this section, we have shown that the  $N$-Lorentz-Weyl transformations of 4-dimensional spacetime found by this method are the familiar rotations, boosts and scale transformations of 4-d spacetime.

\appendix

\section{The N-squared Rep and Basis Reps are Related} \label{Proof}

In this appendix, we prove that there exists a similarity transformation $S$ satisfying (\ref{JN2}), $SJ_{(N^2)}$ = $(h_{(N)} + \bar{h}_{(\bar{N})})S,$ and (\ref{KN2}), $SK_{(N^2)}$ = $-i(h_{(N)} - \bar{h}_{(\bar{N})})S.$ First we prove the theorem for the utility rep in Sec. \ref{UNspacetime} and then extend the result to general reps by considering the transformation from one basis to another.

For the utility rep we show that the similarity transformation $S^{\lambda}_{\sigma}$ is just a reorganization of the basis matrices $h^{\sigma}_{(N)\, mn}.$ The similarity transformation $S^{\lambda}_{\sigma}$ has two indices each running from 1 to $N^2,$ $\lambda,\sigma \in$ $\{1,...,N^2\},$ while $h^{\sigma}_{mn}$ has indices $m$ and $n$ each running from 1 to $N.$ If we make one index $\lambda$ out of the two indices, e.g. $\lambda$ = $N(m-1)+n,$ then we can have $S^{\lambda}_{\sigma}$ =  $h^{\sigma}_{(N)\,mn}$ = $h^{\sigma}_{(N)\, \lambda}.$ 

Start with the fundamental commutation and anticommutation relations (\ref{Comm1}),
$$ [h^{\mu}_{(N)},h^{\nu}_{(N)}]_{mn}  = i f^{\mu \nu \sigma} h^{\sigma}_{(N) \, mn}  \quad ; \quad
 \{h^{\mu}_{(N)},h^{\nu}_{(N)}\}_{mn}  =  d^{\mu \nu \sigma} h^{\sigma}_{(N)\, mn} \quad . \quad {\mathrm{[Eqn. \; (\ref{Comm1})]}} 
$$
The left sides (lhs) may be expanded using the fact that the anti-rep matrices are the negative transpose of the $h_{(N)}$s, i.e. ${\bar{h}}^{\sigma}_{(\bar{N})\, mn}$ = $-h^{\sigma}_{(N)\, nm}.$ One finds that the left side of the commutation relation is
$$ [h^{\mu}_{(N)},h^{\nu}_{(N)}]_{mn}  =  h^{\mu}_{(N)\,ms}h^{\nu}_{(N)\,sn} - h^{\nu}_{(N)\,mt}h^{\mu}_{(N)\,tn} = 
\left(h^{\mu}_{(N)\,ms}\delta^{t}_{n} + \delta^{s}_{m}{\bar{h}}^{\mu}_{(\bar{N})\,nt}\right) h^{\nu}_{(N)\,st} \quad , \quad   {\mathrm{(lhs)}}
$$
By the definitions of $j_{(N^2)}$ in (\ref{jKN2}), the right side is 
 \begin{equation} \label{34left} i f^{\mu \nu \sigma} h^{\sigma}_{(N) \, mn} = i f^{\sigma \mu \nu} h^{\sigma}_{(N) \, mn} = h^{\sigma}_{(N) \, mn} i f^{\sigma \mu \nu} =  h^{\sigma}_{(N) \, mn} j^{\mu}_{(N^2)\sigma \nu}	\quad ,	\quad \quad {\mathrm{(rhs)}}	
\end{equation}
where we use the fact that the coefficients $f^{\mu \nu \sigma}$ are antisymmetric in $\mu \nu\sigma$ in the utility rep by (\ref{fAntidSymm}).

For the anticommutator, we get 
$$ \{h^{\mu}_{(N)},h^{\nu}_{(N)}\}_{mn}   =  h^{\mu}_{(N)\,ms}h^{\nu}_{(N)\,sn} + h^{\nu}_{(N)\,mt}h^{\mu}_{(N)\,tn} = 
\left(h^{\mu}_{(N)\,ms}\delta^{t}_{n} - \delta^{s}_{m}{\bar{h}}^{\mu}_{(\bar{N})\,nt}\right) h^{\nu}_{(N)\,st} \quad , \quad   {\mathrm{(lhs)}}
$$
and, by the definitions $k_{(N^2)}$ in (\ref{jKN2}),
\begin{equation} \label{34right} d^{\mu \nu \sigma} h^{\sigma}_{(N) \, mn} = d^{\sigma \mu \nu} h^{\sigma}_{(N) \, mn} = h^{\sigma}_{(N) \, mn} d^{\sigma \mu \nu} = \epsilon_{P} i h^{\sigma}_{(N) \, mn} k^{\epsilon_{P}\mu}_{(N^2)\sigma \nu}	\quad ,	\quad {\mathrm{(rhs)}}	
\end{equation}
where we use the symmetry of the $d^{\mu \nu \sigma}$s  in $\mu \nu\sigma,$ see (\ref{fAntidSymm}).
Recall that we write $j$ and $k$ in the utility rep and we denote the matrices as $J$ and $K$ for general reps.

Using the new expressions for the left and right sides of (\ref{Comm1}), we get after switching sides,
\begin{equation} \label{34leftrighta} h^{\sigma}_{(N) \, mn} j^{\mu}_{(N^2)\sigma \nu} = \left(h^{\mu}_{(N)\,ms}\delta^{t}_{n} + \delta^{s}_{m}{\bar{h}}^{\mu}_{(\bar{N})\,nt}\right) h^{\nu}_{(N)\,st} \quad , \quad	 \quad {\mathrm{(rhs)=(lhs)}} 
\end{equation}
for the commutation relation and
\begin{equation} \label{34leftright} \epsilon_{P} i h^{\sigma}_{(N) \, mn} k^{\epsilon_{P}\mu}_{(N^2)\sigma \nu} = \left(h^{\mu}_{(N)\,ms}\delta^{t}_{n} - \delta^{s}_{m}{\bar{h}}^{\mu}_{(\bar{N})\,nt}\right) h^{\nu}_{(N)\,st} \quad ,  \quad	 \quad {\mathrm{(rhs)=(lhs)}}
\end{equation}
for the anticommutation relation. Now, with 
\begin{equation} \label{Sj} m, \, n \rightarrow \lambda_{1}, \, \lambda_{0}  \quad ; \quad S^{\lambda}_{\sigma} =   h^{\sigma}_{(N)\,mn} \quad ,
\end{equation}
 where $\lambda$ = $N(m-1)+n$ = $N(\lambda_{1}-1)+\lambda_{0},$ we rewrite (\ref{34leftrighta}) and (\ref{34leftright}) to read
$$ S^{\lambda}_{\sigma} j^{\mu}_{(N^2)\sigma \nu} = \left(h^{\mu}_{(N)\,\lambda_{1} \rho_{1}}\delta^{\rho_{0}}_{\lambda_{0}} + \delta^{\rho_{1}}_{\lambda_{1}}{\bar{h}}^{\mu}_{(\bar{N})\,\lambda_{0}\rho_{0}}\right) S^{\rho}_{\nu} \quad ;	 $$
\begin{equation} \label{Simjk}  S^{\lambda}_{\sigma} k^{\epsilon_{P}\mu}_{(N^2)\sigma \nu} = -\epsilon_{P} i \left(h^{\mu}_{(N)\,\lambda_{1} \rho_{1}}\delta^{\rho_{0}}_{\lambda_{0}} - \delta^{\rho_{1}}_{\lambda_{1}}{\bar{h}}^{\mu}_{(\bar{N})\,\lambda_{0}\rho_{0}}\right) S^{\rho}_{\nu} \quad , 
\end{equation}
where $\rho$ = $N(s-1)+t$ = $N(\rho_{1}-1)+\rho_{0}.$ Thus the theorem is shown for the utility rep.

The utility basis $h^{\mu}_{(N)},$ (\ref{AJN}), forms a basis of all hermitian $N\times N$ matrices. Now consider another basis $h^{\prime \, \nu}_{(N)}$  of hermitian $N\times N$ matrices, where $\mu, \nu \in$ $\{1,...,N^2\}.$ Since $h^{\mu}_{(N)}$ form a basis, it follows that there exists an $N^2 \times N^2$ matrix $R$ with real components such that
\begin{equation} \label{RJ}
h^{\prime \, \mu}_{(N)} = R^{\mu}_{\sigma}h^{\sigma}_{(N)}\quad.
\end{equation}
By the orthonormal property in the utility rep, (\ref{trace1}), we have $R^{\mu}_{\lambda}$ = $2{\mathrm{tr}}(h^{\prime \, \mu}_{(N)}h^{\lambda}_{(N)}).$ The primed basis is a basis for the unprimed $h^{\mu}_{(N)}$s, so $R$ has an inverse, $R^{-1}.$ 

Applied to the anti-rep, the fact that $R$ is real implies that 
\begin{equation} \label{JbarPRIME}\bar{h}^{\prime \, \mu}_{(\bar{N})\, st} = -{h}^{\prime \, \mu \, \ast}_{(N) st} = -\left(R^{\mu}_{\sigma}{h}^{\sigma}_{(\bar{N})\,st}\right)^{\ast} = R^{\mu}_{\sigma}\bar{h}^{\sigma}_{(\bar{N})\,st} \quad ,
\end{equation}
so the anti-rep $\bar{h}$s transform just like the $h$s.

Consider (\ref{Comm1}), i.e. the commutators $[h^{\mu}_{(N)},h^{\nu}_{(N)}]$ = $if^{\mu \nu \sigma}h^{\sigma}_{(N)}$ and anticommutators \linebreak $\{h^{\mu}_{(N)},h^{\nu}_{(N)}\}$ = $d^{\mu \nu \sigma}h^{\sigma}_{(N)}.$ By (\ref{RJ}), the commutators and anticommutators of the primed basis have coefficients 
\begin{equation} \label{fdprime}
f^{\prime \, \mu \nu \lambda} = R^{\mu}_{\sigma}R^{\nu}_{\rho}f^{\sigma \rho \tau}R^{-1 \, \tau}_{\lambda} \quad {\mathrm{and}} \quad d^{\, \prime \, \mu \nu \lambda} = R^{\mu}_{\sigma}R^{\nu}_{\rho}d^{\sigma \rho \tau}R^{-1 \, \tau}_{\lambda} \quad.
\end{equation}
Since $if^{\prime}$ and $id^{\prime}$ are $J^{\prime}_{(N^2)}$ and $K^{\prime}_{(N^2)}$ within a sign, it follows that 
\begin{equation} \label{fdprime1}
{J^{\prime \, \nu}_{(N^2)\, \mu \lambda}} = R^{\mu}_{\sigma}R^{\nu}_{\rho}{j^{ \rho}_{(N^2)\, \sigma \tau}}R^{-1 \, \tau}_{\lambda} \quad {\mathrm{and}} \quad {K^{\epsilon_{P} \, \prime \, \nu}_{(N^2)\, \mu \lambda}} = R^{\mu}_{\sigma}R^{\nu}_{\rho}{k^{\epsilon_{P} \, \rho}_{(N^2)\, \sigma \tau}}R^{-1 \, \tau}_{\lambda} \quad.
\end{equation}
And, lastly, we define $S^{\prime \, \mu}_{\nu}$ 
\begin{equation} \label{fdprime2}
S^{\prime \, \mu}_{\nu} \equiv S^{\mu}_{\sigma} R^{-1 \, \sigma}_{\nu} \quad
\end{equation}
so that, by (\ref{Simjk}) to (\ref{fdprime2}), we have
\begin{equation} \label{JN2appPRIME1}
	S^{\prime \, \lambda}_{\sigma}{J^{\prime \, \mu}_{(N^2)\, \sigma \nu}} = \left(h^{\prime \, \mu}_{(N)\lambda_{1} \phi_{1}}\delta^{\lambda_{0}}_{\phi_{0}} + \delta^{\lambda_{1}}_{\phi_{1}}
{ {\bar{h}^{\prime \, \mu}} }_{(\bar{N})\lambda_{0} \phi_{0}} \right) S^{\prime \, \phi}_{\nu} \quad
\end{equation}

\begin{equation} \label{KN2appPRIME1}
S^{\prime \, \lambda}_{\sigma}K^{(\epsilon_{P}) \, \prime \, \mu}_{(N^2)\, \sigma \nu} = -\epsilon_{P} i \left(h^{\prime \, \mu}_{(N)\lambda_{1} \phi_{1}}\delta^{\lambda_{0}}_{\phi_{0}} - \delta^{\lambda_{1}}_{\phi_{1}}
{ {\bar{h}^{\prime \, \mu}} }_{(\bar{N})\lambda_{0} \phi_{0}} \right) S^{\prime \, \phi}_{\nu}   \quad,
\end{equation}
where $\lambda$ = $N(\lambda_{1}-1)+\lambda_{0}$ and $\phi$ = $N(\phi_{1}-1)+\phi_{0}.$  Removing the primes, we have shown that there is a similarity transformation $S$ satisfying (\ref{JN2}), $SJ_{(N^2)}$ = $(h_{(N)} + \bar{h}_{(\bar{N})})S,$ and (\ref{KN2}), $SK_{(N^2)}$ = $-i(h_{(N)} - \bar{h}_{(\bar{N})})S$ for a general basis $h_{(N)}$ and the generators $J_{(N^2)}$ and $K_{(N^2)}$ of its associated $N^2$-dimensional $N$-Lorentz-Weyl rep. End-of-proof.


\end{document}